\documentclass[journal=nalefd,manuscript=article]{achemso}

\usepackage[version=3]{mhchem} % Formula subscripts using \ce{}
\usepackage{achemso}
\author{Bruno Cury Camargo}
\affiliation{Institute of Physics, Polish Academy of Sciences, Aleja Lotnikow 32/46, PL-02-668 Warsaw, Poland}
\alsoaffiliation{Laboratoire National des Champs Magnétiques Intenses, CNRS-INSA-UJF-UPS, UPR3228; 143 avenue de Rangueil, F-31400 Toulouse, France.}
\email{b.c_camargo@yahoo.com.br}
\author{Benjamin Lassagne}
\affiliation{LPCNO, Université de Toulouse, CNRS, INSA, UPS, 135 avenue de Rangueil, 31077 Toulouse, France}
\author{Raul Arenal}
\affiliation{Laboratorio de Microscopias Avanzadas (LMA), Instituto de Nanociencia de Aragon (INA), U. Zaragoza, C/ Mariano Esquillor s/n, 50018 Zaragoza (Spain).}
\altaffiliation{Fundacion ARAID, 50018 Zaragoza, Spain.}
\author{Christophe Gatel}
\affiliation{Centre d’Elaboration de Matériaux et d’Etudes Structurales, CNRS, 29 rue Jeanne Marvig, F-31077 Toulouse (France)}
\author{Thomas Blon}
\author{Guillaume Viau}
\author{Lise-Marie Lacroix}
\email{lmlacroi@insa-toulouse.fr}
\affiliation{LPCNO, Université de Toulouse, CNRS, INSA, UPS, 135 avenue de Rangueil, 31077 Toulouse, France}
\author{Walter Escoffier}
\affiliation{Laboratoire National des Champs Magnétiques Intenses, CNRS-INSA-UJF-UPS, UPR3228; 143 avenue de Rangueil, F-31400 Toulouse, France.}

\title{Platinum tripods as nanometric frequency multiplexing devices.}


\begin{document}
%%%%%%%%%%%%%%%%%%%%%%%%%%%%%%%%%%%%%%%%%%%%%%%%%%%%%%%%%%%%%%%%%%%%%
%% The manuscript does not need to include \maketitle, which is
%% executed automatically.  The document should begin with an
%% abstract, if appropriate.  If one is given and should not be, the
%% contents will be gobbled.
%%%%%%%%%%%%%%%%%%%%%%%%%%%%%%%%%%%%%%%%%%%%%%%%%%%%%%%%%%%%%%%%%%%%%
\begin{abstract}
Electrical and structural characterization of nano-particles are very important steps to determine their potential applications in microelectronics. In this paper, we adress the crystallographic and electric transport properties of soft-chemistry-grown nanometric Pt tribranches. We report that Pt nanostars grown from the reduction of $\text{H}_{2}\text{PtCl}_6$ salt in pure oleylamine present a remarkable crystalline structure and a deeply metallic character despite being grown under mild conditions. We demonstrate that such devices are able to operate at current densities surpassing $200$ MA/$\text{cm}^2$, actuating as highly compact frequency multiplexers in the non-ohmic regime.
\end{abstract}

\section*{Introduction}
Nowadays, a progressive miniaturization of electrical devices to the nanometric scale is imperative for achieving more densely-packed processing power. This calls for carefully engineered components at the scale of a few tens of nanometers. Such an extreme miniaturization constitutes a major manufacturing challenge, as surface and short-range effects (such as van-der-Waals, dipolar interaction, electrostatic, etc…) add complexity to the design and fabrication processes compared to micrometer-scale devices.

A possible route for miniaturization is the bottom-up approach, where the building blocks are nano-objects grown using chemical methods. Recent developments in nanochemistry yields high-quality nano-crystals with tuneable sizes and shapes.\cite{Kwon2008, Tao2008} Such liquid-phase syntheses are generally governed by successive nucleation and growth steps. The modulation of the precursors reactivity, by thermal or chemical means, constitutes one of the major levels of control of the final nanoparticle (NP),\cite{Shevchenko2002} leading to an unprecedented shape control. 

Among the different materials grown using chemical methods, noble metal NPs have been extensively studied in the past few years. For example, kinetically-controlled chemical growth conditions can lead to the formation of Au,\cite{Park2015} Pd,\cite{Watt2009, Chu2012} Pt,\cite{Maksimuk2006, Maksimuk2007} and bimetallic stars.\cite{Zhang2016}In particular, Maksimuk et al. reported that a two-step nucleation/growth process could lead to Pt tripods.\cite{Maksimuk2007} Elaborating on this possibility, we subsequentially demonstrated, in a previous work, a reproducible achievement of single-crystalline 3-fold (tripods) and 5-fold Pt stars by tuning the decomposition of $\text{H}_{2}\text{PtCl}_6$ in presence of dihydrogene.\cite{Lacroix2012} These objects could be model systems for electronic transport measurement due to their perfect crystallinity. 

Yet, the electric and electronic properties of such star-shaped metallic NPs did not receive much attention to date, with reports in the literature mostly focusing on their plasmonic and catalytic properties.\cite{Guerrero2011, Yin2011} However, because of their high crystallinity and multi-branch nature, these  are foreseen as active or interconnecting parts of the next generation of miniaturized circuits. For instance, their semiconducting counterparts were reported to selectively deflect an incoming electron beam from the central branch towards the second or third branch thanks to a local electric field generated by side-gate electrodes.\cite{Hieke2000} 

These perspectives are just being envisioned at the moment as more fundamental studies are required. In order to improve the understanding of such exotic objects, in this paper, we experimentally determine the electric properties of chemically-grown, highly crystalline, star-shaped platinum nanostars. Our samples have shown the properties of bulk metals, thus operating as classical devices. Thanks to their geometry, these objects show non-linear electrical properties with potential application as nanometric, passive frequency multipliers.

\section*{Results and discussion}

\begin{figure}[h]
  \includegraphics[width=12cm]{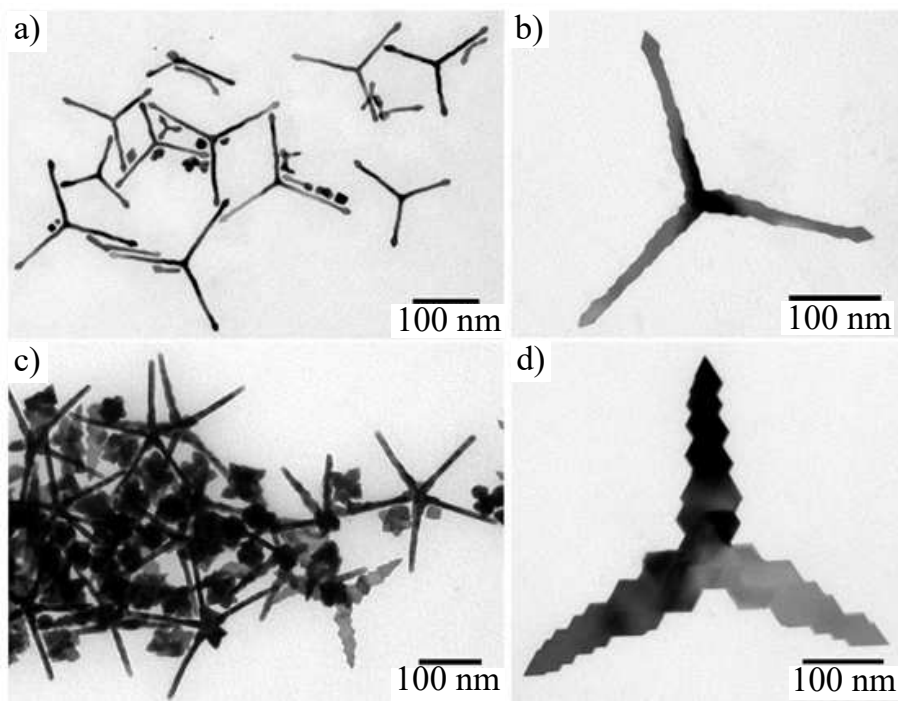}
  \caption{Transmission electron microscopy images of Pt nanostars obtained after $48$h of reduction under $\text{H}_2$ of $\text{H}_2\text{PtCl}_6$ a-b) As purchased. c-d) After $48$h of aging  the starting solution of $\text{H}_2\text{PtCl}_6$ in oleylamine at $150^o$C under air. b) and d) High magnification image of 3-fold stars. }
  \label{fig_stars}
\end{figure}

Pt nanostars were obtained by reducing a platinum salt ($\text{H}_2\text{PtCl}_6$, $\text{6H}_2\text{O}$) in pure oleylamine under dihydrogene at $150^o$C, as previously described.\cite{Lacroix2012} The direct decomposition of the precursor at low concentration (here typically $2.5$ mM)  yield thin nanostars along with few ill-defined nanoparticles. The stars exhibit mostly a three-fold symmetry with mean arm width and length of $7\pm 1$ nm and $60\pm 35$ nm, respectively (Figure \ref{fig_stars} and S1). The dimension of the nanostars was controlled by adjusting the precursor reactivity. An intermediate Pt-oleylamine complex was obtained by aging $10$ mM of $\text{H}_2\text{PtCl}_6$ in pure oleylamine at $150^o$C under air for $48$h. Thick nanostars, exhibiting a three-fold or a five-fold symmetry, could be obtained by decomposing this intermediate precursor under $\text{H}_2$ for another $48$h. The tripods dimension were much larger compared to the direct reduction, with a mean arm width and length of $52\pm 21$ nm and $230\pm 70$ nm respectively (Figure S2).  Interestingly, these stars exhibit numerous sharp edges, as previously described.\cite{Maksimuk2007, Maksimuk2006}

At shorter reaction time, decahedra and triangular platelets were observed (Figure S3-S4). The latter acted as seeds for the further overgrowth of the three arms. By adjusting the reactivity of the precursor, varying from $\text{H}_2\text{PtCl}_6$ salt to an oleylamine-Pt complex, we could finely tune the initial seeds’ size from $7$ and $50$ nm. While other experimental parameters such as the reaction time and the Pt concentration were previously used to tune the arms' dimensions,\cite{Chandni2011} the aging of the platinum salt in presence of oleylamine was found to be less subject to reproducibility issues.  

\begin{figure}[h]
  \includegraphics[width=12cm]{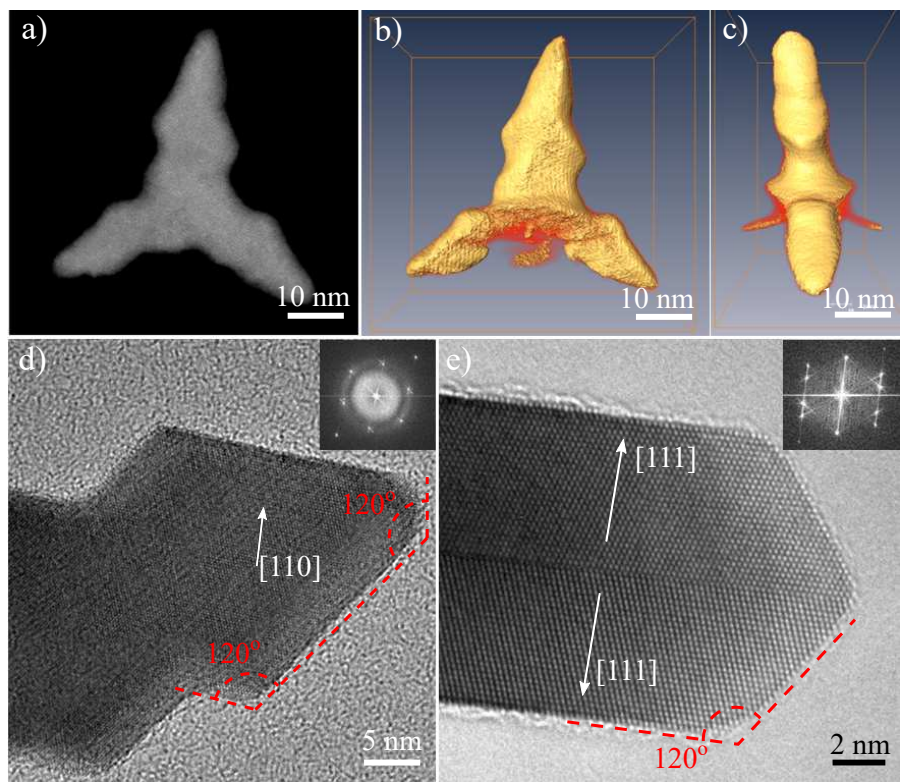}
  \caption{a-c) High Angle Annular Dark Field TEM image and the corresponding $3$D reconstruction obtained by electron tomography with (b) top and (c) lateral view. d-e) High resolution TEM images of thick 3-fold Pt stars from a top-view (d) and lateral view (e). The twinning plane is clearly visible. The insets in d) and e) show the corresponding FFT images. }
  \label{fig_tomo}
\end{figure}

Electron tomography of an isolated nanostar has been performed in order to gain deeper understanding of its peculiar shape (video S1). Such experimental technique allows for a complete $3$D reconstruction of the object based on a series of $2$D TEM images acquired by High Angle Annular Dark Field Scanning TEM at different tilting angles. In the star shown in fig. \ref{fig_tomo}, despite a large missing wedge in the electron tomography series, the structure of the upper branch could be resolved. Results revealed a star with a flat surface and a thickness of $7.5$ nm. 

High resolution TEM images performed on thick stars confirmed their good crystallinity (Figure \ref{fig_tomo}). The presence of tilted branches allowed to clearly observe a twin plane, separating the arms into two mirror-like perfect crystals, as previously reported.\cite{Maksimuk2007, Lacroix2012}  The overgrowth of the Pt arms keept on propagating the twin plane, initially present within the triangular seeds (Figure S4). The presence of the sharp edges and the twin planes leads to a star exhibiting mostly \{111\} facets, which are known to have the lowest surface energy in Pt. 

Planar 3-fold stars have been further studied thanks to proper electric connection. The devices were divided into two categories: thick and thin nanostars.  Each arm in the thick samples had typical dimensions of $250$ nm x $60$ nm x $16$ nm (length x width x height), whereas the thinner Pt nanostars had typical dimensions of $100$ nm x $7$ nm x $7$ nm (see the suppl. material for an AFM image of a thick device). In both cases the nanostars were composed by two perfect crystals grown symmetrically aside a twinning boundary (Figure \ref{fig_tomo}e). 

\begin{figure}
  \includegraphics[width=12cm]{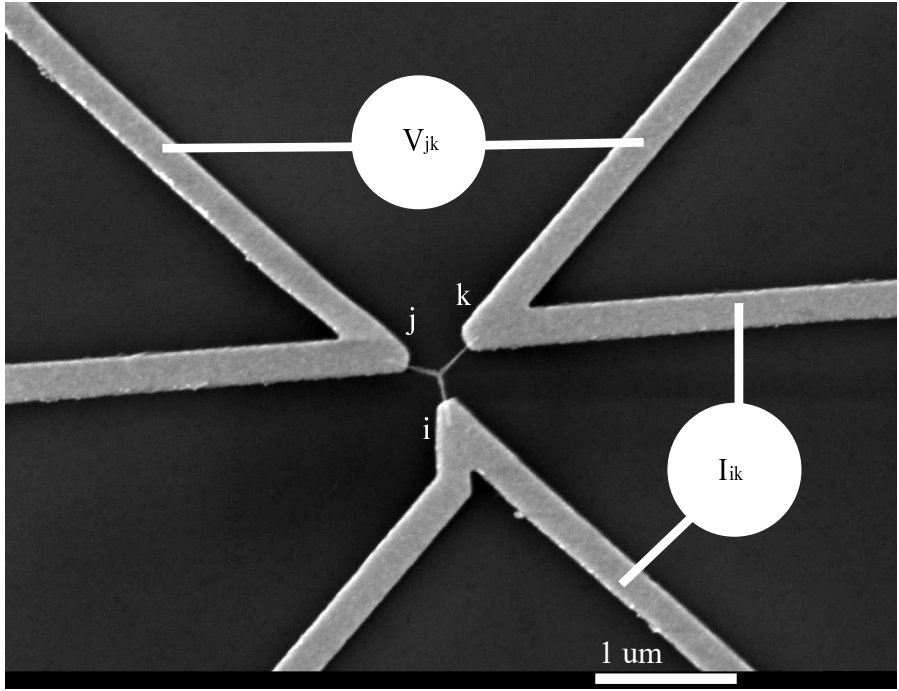}
  \caption{Scanning electron microscopy image of a contacted thin Pt nanostar. The letters $i$, $j$, $k$ are used to identify each branch.}
  \label{fig_contacted}
\end{figure}

The samples were electrically addressed as described in the methods section. In short, a solution containing the stars was deposited atop $300$ nm $\text{SiO}_2$-coated Si substrates. After cleaning processes, the cleanest and most isolated specimens were selected and contacted using standard electron nanolithography techniques. The electrodes were composed by a thermally-evaporated $40$ nm gold film on a $5$ nm Chromium buffer layer. A connected device is shown in figure \ref{fig_contacted}, where the branches are labeled with indexes $i$, $j$ and $k$. Measurements were performed in the three-probe measurement scheme, in which an electrical current is injected through the branches $i$ and $k$ ($I_{ik}$), while the voltage ($V_{jk}$) is measured between branches $j$ and $k$. The ratio $V_{jk}/I_{ik}$ provides the resistance of branch $k$, denoted $R_k$. This experimental resistance can be viewed as the upper limit of the star's dendrite resistance, which includes an additional contact resistance (see the supplementary material for more details).

\begin{figure}
  \includegraphics[width=12cm]{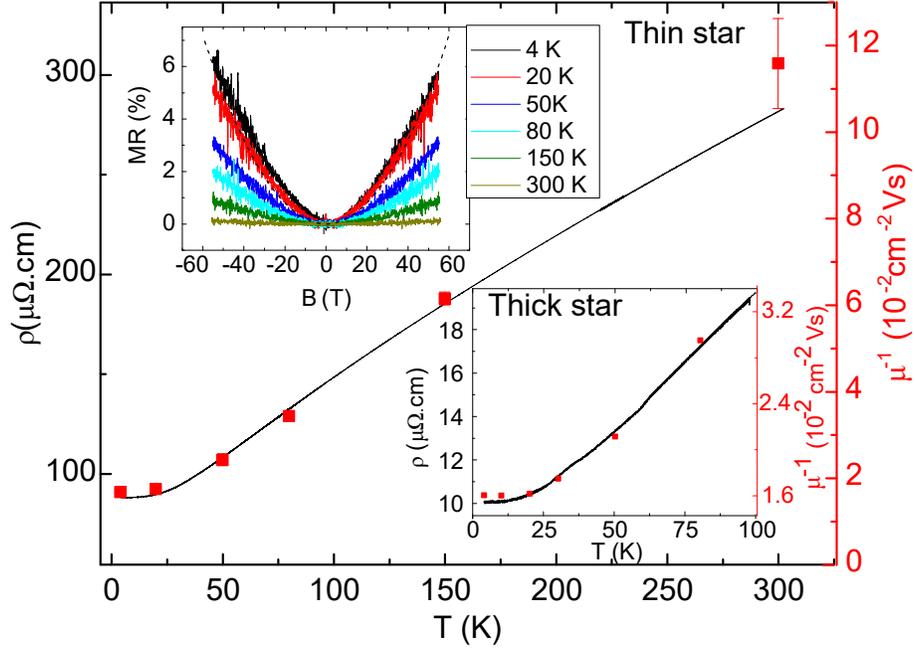}
  \caption{Main panel: Electrical resistivity (black line, left axis) and inverse of electronic mobility (closed points, right axis) for the thin nanostar. Lower inset: The same as the main panel, but for a thick nanostar. Upper inset: magnetoresistance curves measured for the thin nanostar at selected temperatures. The curves at $\text{T} = 20$ K, $50$ K and $150$ K were mirrored to ease visualization. The dashed line is a function of the type $\text{MR} = \mu^{2} B^2$ with $\mu \approx 44 \text{ cm}^2\text{V}^{-1}\text{s}^{-1} = 4.4\times10^{-3}\text{ T}^{-1}$.}
  \label{fig_RxT}
\end{figure}

All branches of the same star presented similar properties, and hence only a single branch of each sample will be presented. The typical temperature dependency of the branches electrical resistivity is shown in figure \ref{fig_RxT}. The stars showed a behavior of the type $\rho (T)-\rho (T=0) \propto T^{5/2}$ at low temperatures, transitioning to $\rho(T)\propto T$ above $100$ K. This behavior is typical of metals and is usually attributed to thermally-activated electron-phonon scattering.\cite{Ashcroft} Measurements at magnetic fields up to $60$ T have shown a quadratic magnetoresistance (MR), 
\begin{equation}
\text{MR} \equiv \frac{R(B)-R(B=0)}{R(B=0)} =\mu^{2} B^{2},
\end{equation} 
which remained below $10$\%. The term $\mu$ is obtained from the fitting of the experimental data and is assigned to the Drude-like mobility of the system.\cite{Pippard} Values obtained this way were in the range $8 \text{ cm}^2\text{V}^{-1}\text{s}^{-1}<\mu <70 \text{ cm}^2\text{V}^{-1}\text{s}^{-1}$ (in practical units). The inverse mobility ($\mu^{-1}$) scaled together with $\rho(T)$, suggesting the reduction of $\mu$ as the main mechanism for the $\rho(T)$ behavior shown. Such dependence is expected in bulk metals in the diffusive regime according to the Drude model ($\sigma=\rho^{-1}=ne\mu$),\cite{Pippard} and indicates that the samples, albeit nanometric, behave as well-crystallized bulk metals.

Our results are consistent with previous studies carried out in high-purity Pt. For example, it was experimentally inferred that the mean free path for electrons in pure Pt thin films lies around $10$ nm, with finite size effects being more pronounced in samples with thicknesses  below $1$ nm.\cite{Fischer1980} At the same time, ab-initio calculations estimate an inelastic mean free path below $5$ nm for electrons in Pt.\cite{Fischer1980, Zhukov2006} These values are smaller than the typical dimensions of our devices. In addition, the residual resistivity ratio in our samples ($\text{RRR}\equiv \text{R}(300\text{ K})/\text{R}(4.2 \text{ K})$) was below $10$, suggesting the presence of a relatively high concentration of impurities (typical RRR's in high-purity bulk Pt range between $\text{RRR} \approx 50 - 5000$ \cite{Tew2013, Shah1972, McGee}), and thus a shorter mean free path than the ones reported in the literature.\cite{Fischer1980, Zhukov2006}  However, we note that the amount of impurities in our devices did not greatly increase their expected resistivity, which ranged between $100\text{ } \mu \Omega \text{cm} < \rho (300\text{ K})<300 \text{ } \mu \Omega \text{cm}$. Such values are only one order of magnitude higher than the typical resistivity of pure Pt (approx. $10$ $\mu \Omega$cm),\cite{McGee, Giancolli} thus corroborating that impurities/defects in our samples do not mask the metallic character of the crystalline NPs.

\begin{figure}
  \includegraphics[width=12cm]{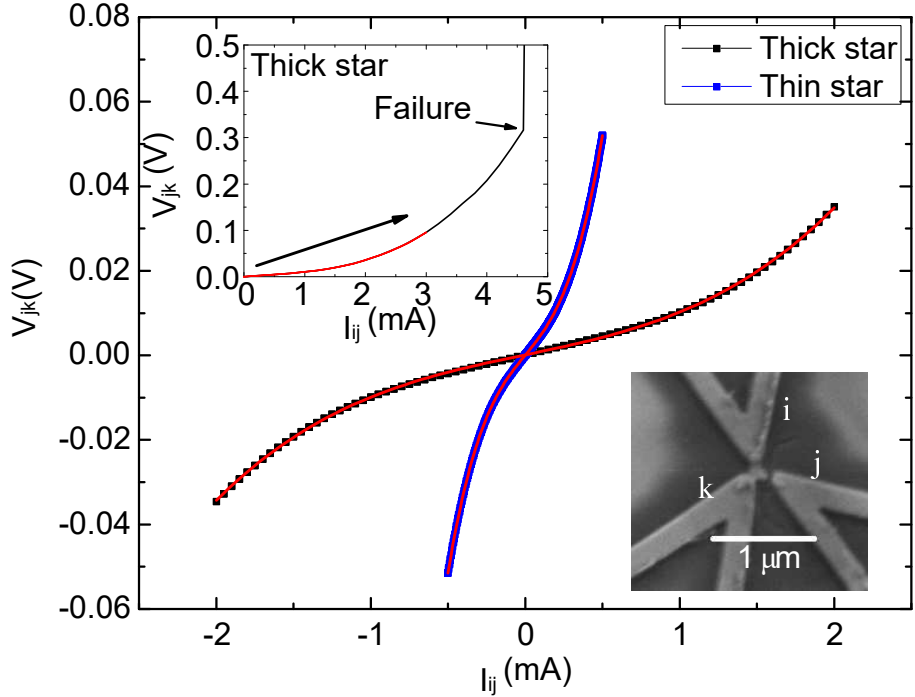}
  \caption{I(V) curves for a thin (blue points) and thick (black points) stars at $\text{T} = 4.2$ K. The red lines are fits of the type $\text{V}(\text{I})=\text{I}\times(\text{R}_0+\alpha \text{I}^2)$, with $\text{R}_0=44.68(1)$  $\Omega$ and $\alpha=2.35(1)\times10^8$  $\Omega/\text{A}^2$ for the thin sample, and with $\text{R}_0=7.66(1)$  $\Omega$ and $\alpha=2.410(1)\times 10^6$  $\Omega/\text{A}^2$ for the thick sample. %{The dashed lines are fits of the type $\text{I}(\text{V})=\text{V}/(\text{R}_0+\beta \text{V}^2+\gamma \text{V}^4)$, with $\text{R}_0=8.82(3)$  $\Omega$, $\beta=12.4(1)\times 10^3$ $\Omega / \text{V}^2$ , and $\gamma=4.1(1)\times 10^6$ $\Omega / \text{V}^4$ for the thick star, whereas $\text{R}_0=50.8(1)$  $\Omega$, $\beta=37.1(2)\times 10^3$  $\Omega / \text{V}^2$, and $\gamma=6.9(1)\times 10^6$  $\Omega / \text{V}^4$ for the thin sample.}
The top inset shows the destructive IxV test for a thick star to assess its maximum supported electrical current. The lower inset shows a picture of the destroyed sample.}
  \label{fig_IxV}
\end{figure}

Current versus voltage characteristics of the devices are shown in fig. \ref{fig_IxV}. At applied currents $\text{I}>500$ $\mu$A, all samples departed from the ohmic behavior. For much larger electrical excitations, the devices did fail due to excessive heating. The failure took place at the nanostars branches, rather than at the contacts, thus showing the superior electrical resistance of the former over the latter. The electrical current necessary for such failure was of $3 - 5$ mA (sample dependent), corresponding to an average electric current density above $200$ $\text{MA}/\text{cm}^2$. For reference, this amount is two orders of magnitude above typical values in conventional semiconducting integrated devices.\cite{Lienig2013} 

\begin{figure}

\centering
	%\begin{subfigure}
		\includegraphics[width=12cm]{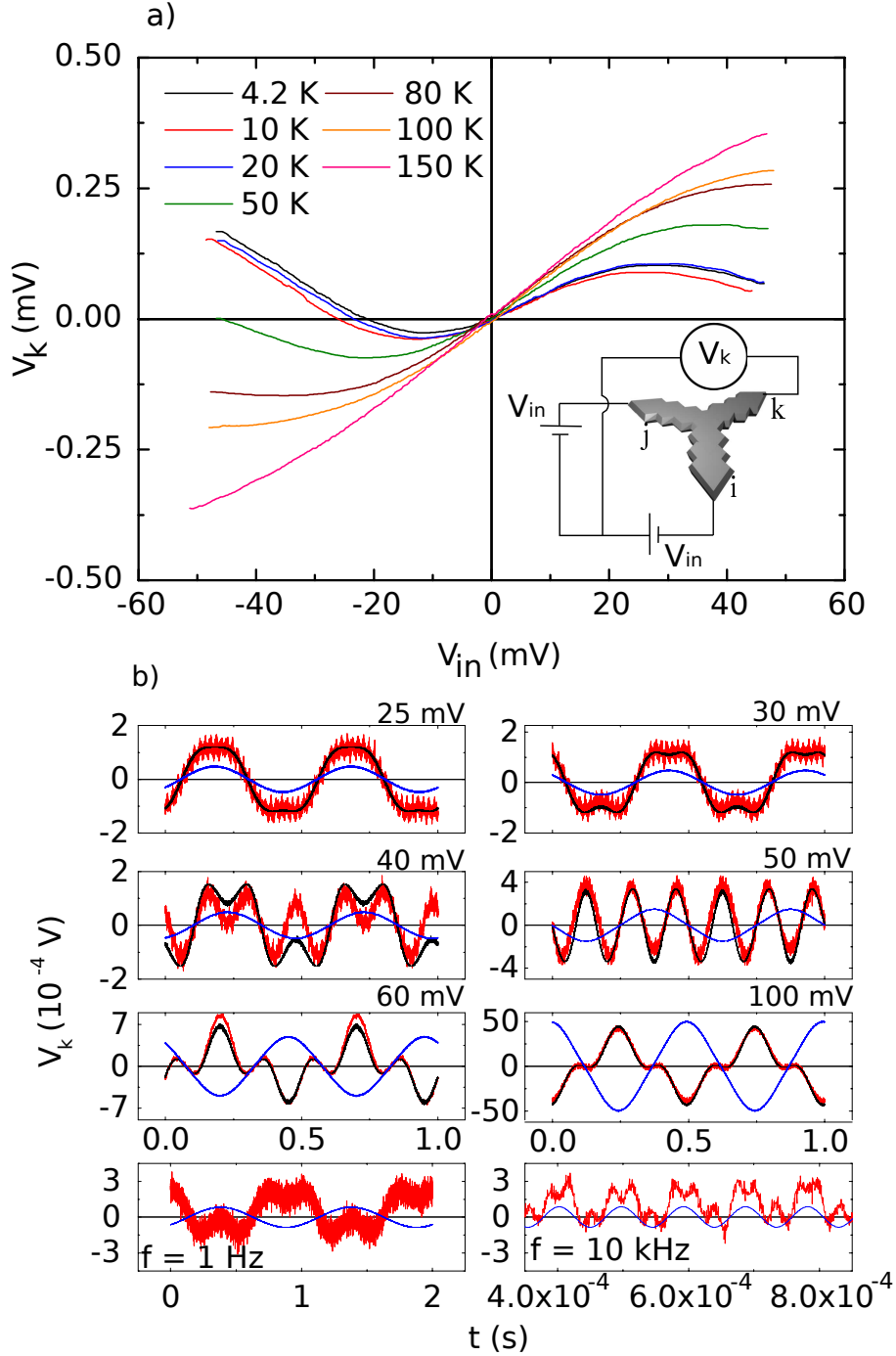}
	%\end{subfigure}
	%\begin{subfigure}
	%
	%	\includegraphics[width=10cm]{figure_5b.eps}
	%\end{subfigure}
\caption{a) Smoothed $\text{V}_k(\text{V}_{in})$ in the push-pull configuration for a thin star at different temperatures (see the suppl material for the raw data). The inset shows how the measurement was performed.  b) $\text{V}_k$ for different values of excitation voltage $\text{V}_{in}$ and at two different frequencies at $\text{T} = 4.2$ K. The blue curves correspond to Vin (not in scale with the data). The red curves correspond to the experimental data, and the black curves are obtained from $\text{V}_k=(\text{V}_{in} \delta_0+\text{V}_{in}^3 \delta_\beta)/(\Delta_0 + \text{V}_{in}^2 \Delta_\beta)$, using $\delta_0 = 0.1755$ $\Omega$, $\delta_\beta=243$ $\Omega /V^2$, $\Delta_0=15$ $\Omega$ and $\Delta_\beta=20000$ $\Omega /V^2$. The same parameters were used for all curves. Note the frequency doubling and tripling taking place at $\text{V}_{in} = 40$ mV and $\text{V}_{in} = 50$ mV, respectively.}
  \label{fig_freq}
\end{figure}

Prior to failure, the star’s resistance was well described by the empirical relation $\text{R}_j (\text{I}_{ij} )=(\text{R}_{0_j} +\alpha_j \text{I}_{ij}^2 )$ where the non-linear term can be understood as due to a variation of the device temperature because of Joule heating. Thanks to the unusual geometry of the stars, we were able to explore the implications of such non-linearity by measuring the samples in a configuration known as push-pull. In it, symmetric voltage sources are used to simultaneously drive two different sample branches, as depicted in fig. \ref{fig_freq}a. Classically, the voltage at the third branch relates to the difference between the resistances of the remaining branches as:
\begin{equation}
\text{V}_k=\text{V}_{in}(\text{R}_j-\text{R}_i)/(\text{R}_j+\text{R}_i),
\label{eq_Vk}
\end{equation}
where $\text{V}_{in}$ corresponds to the input voltage (see the suppl. material for the equivalent circuit). Expanding each $\text{R}_j(\text{V}_{in})=R_{0_j}+\beta_j\text{V}_{in}^2+\mathcal{O}(V_{in}^4)$, keeping the terms up to the second order in V and substituting it in eq. \ref{eq_Vk} yields an expression of the type $\text{V}_k=(\text{V}_{in} \delta_0+\text{V}_{in}^3 \delta_\beta)/(\Delta_0 +\text{V}_{in}^2 \Delta_\beta)$, with $\delta_0 = \text{R}_{0_j}-\text{R}_{0_i}$, $\delta_\beta = \beta_j-\beta_i$, $\Delta_0 = \text{R}_{0_j}+\text{R}_{0_i}$ and $\Delta_\beta = \beta_j+\beta_i$ (see the suppl. material for an explicit deduction). We use this geometry to explore the non-linear behavior of the device.

For this purpose, a low $1$ Hz excitation voltage $\text{V}_{in}$  was applied in a push-pull fashion at branches $i$ and $j$ of a thin star, while the voltage at the third branch was recorded. Results are shown in fig. \ref{fig_freq}, where $\text{V}_k$ is plotted for different $\text{V}_{in}$ at $\text{T} = 4.2$ K. As the input voltage increases, the non-linear terms in $\text{V}_k(\text{V}_{in})$ become more important, causing a frequency multiplication. In particular, we observe frequency doubling and tripling effects for $\text{V}_{in} = 40$ mV and $\text{V}_{in} = 50$ mV, respectively. The phenomenon is more pronounced at low temperatures, as shown in fig. \ref{fig_freq}a. We conjecture that this happens because less power is necessary to locally increase the sample temperature at low T, thus resulting in a more pronounced non-ohmicity of the devices at cryogenic temperatures. 

Albeit the non-linearity observed can be attributed to the temperature change of the devices due to joule heating, the sample's reduced dimensions, large surface/volume ratio and metallic character are crucial, as they warrant a system with efficient heat exchange. This is evidenced in $\text{V}_k(\text{V}_{in})$ curves shown in fig. \ref{fig_freq}a, which are reproducible in the non-linear regime upon increasing and decreasing $\text{V}_{in}$. Samples operated successfully in the non-linear regime up to $10$ kHz (the highest $\text{V}_{in}$ frequency probed), further demonstrating the low thermal inertia of the stars (see figure \ref{fig_freq}b).

We note, however, that an unexpected DC rectification effect is seen in some devices. This arises as a slight asymmetry in their $\text{V}_k(\text{V}_{in})$ curves (see fig. \ref{fig_freq}), as well as a non-zero averaged value of $\text{V}_k(t)$. The rectification term can be understood if one considers that the IxV characteristics of the devices are slightly asymmetric, in which case R(I) could be written as $\text{R}=\text{R}_0+\lambda \text{I}+ \alpha \text{I}^2$, $\lambda \ll \text{R}_0 , \alpha$. The $\lambda$ term of each branch becomes increasingly important for devices with very small $\text{R}_j-\text{R}_i$ (see eq. \ref{eq_Vk}). Albeit incompatible with the hypothesis that the samples behave as classical metals, and usually attributed to ballistic effects in semiconducting tribraches,\cite{Shorubalko2001} such behavior can be understood classically if one considers the geometry in which our measurements were made. $\text{V}_k$ measures the difference of resistance not only between branches, but also between their respective contacts. Any non-linearity at the sample contacts, hence, will translate to $\text{V}_k$. Since the contacts are not made from the same material as the Pt device, it is reasonable to assume the occurrence of an intermetallic barrier at the interface, similar to a very leaky diode. Assuming such design, the rectification observed in $\text{V}_k$ is justified within classical phenomena.	

In conclusion, we characterized the electrical properties of highly-crystalline Pt nanostars grown at $150^o$C and using soft chemistry methods.The devices have shown electrical resistivity in the range $100 \text{ } \mu \Omega \text{cm} < \rho(300\text{ K})<300 \text{ } \mu \Omega \text{cm}$ - only one order of magnitude higher than values reported for high-purity, pristine Pt. Such behavior can be attributed to the occurrence of a small quantity of impurities/defects, which are not enough to hinder the highly metallic character of the NPs. Indeed, the nanostars have shown transport characteristics typical of well-crystalized bulk metals, lacking finite-size phenomena despite their reduced dimensions. We have demonstrated that, due to local heating, these devices can operate as highly compact frequency multipliers in the non-linear regime without the need for semiconductors. Their reduced dimensions result in small thermal inertia and cause the frequency multiplication regime to settle for low excitations. At the same time, the sample metallic character allows the devices to withstand current densities in the order of $200$ $\text{MA/cm}^2$, two orders of magnitude above values used in conventional semiconducting systems. Our results open routes towards the growth of cheap, exceedingly high-quality crystals at low temperatures (and low cost) for condensed-matter physics and electrical engineering applications. 

\section*{Methods}

\section*{Synthesis of Pt nanostars.}
Pt nanostars were prepared by reduction of $\text{H}_2\text{PtCl}_6$ salt in pure oleylamine by adapting a previously published synthesis.\cite{Lacroix2012} Briefly, $\text{H}_2\text{PtCl}_6$ (Alfa Aesar, $10$ mg, $0.02$ mmole) and oleylamine (Aldrich, $10$ mL) were mixed in a vial, and placed $15$ mn in an ultrasonic bath to dissolve the Pt precursor. The $2$ mM solution was transferred in a Fisher-Porter bottle and pressurized up to $3$ bars of $\text{H}_2$. The bottle was then let to react at $150^o$C in a pre-heated oil bath for $48$ hour without any stirring. After $48$h, a black precipitate is obtained along with a yellow supernatant, indicating that some Pt precursor still remained. The excess of oleylamine and unreacted species were removed by classical centrifugation process. $40$ mL of hexanes (Aldrich) was added to solubilise the NPs. $40$ mL of absolute ethanol (VWR) was added and the solution was then centrifugated ($4000$ rpm, $6$ min) to separate the NPs. The process is repeated $3$ times. The final product (ca. $5$ mg) was kept in powder form or diluted in $5$ mL Toluene (Carlo Erba) for further use. 

The thickness of the stars could be tuned by aging a $10$ mM solution of $\text{H}_2\text{PtCl}_6$ precursor ($51$ mg, $0.1$ mmole) in oleylamine ($10$ mL) at $150^o$C  for $48$h under an air blanket. The color of the solution slightly changed towards orange. The solution is then pressuriez to $3$ bars of $\text{H}_2$ at room temperature and then allow to react $48$h at $150^o$C. After purification, $25$ mg of black powder is obtained.

\section*{Characterizations}
Characterizations: Samples for transmission electron microscopy (TEM) were prepared by depositing few drops of diluted solution on amorphous carbon coated cupper grid. Low resolution images were obtained with a JEOL-1400 microscope, operating at $120$ kV. High resolution TEM images were obtained with a Tecnai F20 ($200$ kV) equipped with a spherical aberration corrector. Tomography HAADF-HRSTEM measurements were performed on probe-corrected scanning TEM (STEM) FEI Titan Low-Base $60-300$ operating at $300$ keV (fitted with a X-FEG® gun and Cs-probe corrector (CESCOR from CEOS GmbH)). The tilt series were obtained by tilting the specimen in the angular range of $\pm 66^o$ using an increment of $2^o$ in the Saxton mode.\cite{Saxton1984, Book_Deepak} The tilt series data were treated for imaging processing, alignment and reconstruction, using INSPECT 3D (FEI) and the volume reconstruction was performed employing the simultaneous iterative reconstruction technique (SIRT) (100 iterations). The volume segmentation and the visualization were performed using Amira software (FEI). It is worth noting that no evidence of irradiation damage in the samples was detected during the tilt series acquisition.

\section*{Connection of Pt 3-fold stars}
$0.5$ mg of Pt nanostar powder is dispersed into $4$ mL of Toluene by sonication ($5$ min). Two drops of this diluted solution are deposited on a on an insulating oxydized silicon substrate patterned with an alpha-numeric grid. The substrate is then rinsed several times with pure toluene to remove as much contaminants as possible and then dried with a powerful nitrogen blow, in order to prevent its natural evaporation. The sample is later observed with SEM. Isolated nano-stars with the adequate dimensions are selected and localized for individual electrical addressing. The samples are then covered with a $400$ nm thick layer of PMMA resist, which is allowed to cure for $120$ s at $180^o$C. The electron lithography writing is carried out with a dose of $120$ $\mu\text{C/cm}^2$ at acceleration potential of $20$ kV. We use a two-steps process to define the electrodes. First, micro-electrodes are created close to the nanostars together with additional alignment marks. After metal deposition and lift-off of the resist with acetone, the position of the nanostars is accurately determined by SEM relatively to the new alignment marks. A second-step e-beam lithography process is then used to link with precision the arms to the micro-electrodes. To ensure a good electrical contact, the metal deposition is preceded by a soft oxygen plasma etching ($45$ s, $15$ W) to remove any resist residue or ligands.

\begin{acknowledgement}

We thank N. Bruyant and F. Vigneau for enlightening discussions, L. Drigo for technical support and A. Pierrot for assistance with AFM measurements. We also thank the Atelier Interuniversitaire de Micro-nano Electronique of INSA-Toulouse for technical support in electrically addressing our devices. This work has received funding from the People Programme (MarieCurie Actions) of the European Union’s Seventh Framework Programme (FP7/2007-2013)under REA grant agreement n. PCOFUND-GA-2013-609102, through the PRESTIGE programme coordinated by Campus France. It was also supported by the Programme Investissements d'Avenir under the program ANR-11-IDEX-002-02 ref. ANR-10-LABX-0037-NEXT and by the European Union Seventh Framework Program under Grant Agreement 312483-ESTEEM2 (Integrated Infrastructure Initiative-I3). LML, CG and RA gratefully acknowledge the European Associated Laboratory (LEA)−TALEM2 for financial support. High magnetic field measurements were performed at LMCMI under proposal TMS02-215.

\end{acknowledgement}

\section{Supporting Information} Supplementary information file with size distribution and sample characteristics, raw IxV curves and details about the phenomenological model discussed.  Video showing the 3D rendering of a nanostar. 

\providecommand{\latin}[1]{#1}
\makeatletter
\providecommand{\doi}
  {\begingroup\let\do\@makeother\dospecials
  \catcode`\{=1 \catcode`\}=2\doi@aux}
\providecommand{\doi@aux}[1]{\endgroup\texttt{#1}}
\makeatother
\providecommand*\mcitethebibliography{\thebibliography}
\csname @ifundefined\endcsname{endmcitethebibliography}
  {\let\endmcitethebibliography\endthebibliography}{}

%% graphical table of contents.
%%%%%%%%%%%%%%%%%%%%%%%%%%%%%%%%%%%%%%%%%%%%%%%%%%%%%%%%%%%%%%%%%%%%%

\end{document}